\documentclass[aps,twocolumn,amsmath,amssymb,superscriptaddress,prd]{revtex4-1}
\usepackage{graphicx, subfigure}

\begin{document}

\author{Alan O. Jamison}
\email{jamisona@uw.edu}
\affiliation{University of Washington Department of Physics, Seattle, Washington 98195, USA}

\date{\today}

\begin{abstract}
Considering the existence of old neutron stars puts strong limits on the dark matter/nucleon cross section for bosonic asymmetric dark matter. Key to these bounds is formation of a Bose-Einstein condensate (BEC) of the asymmetric dark matter particles. We consider the effects of the host neutron star's gravitational field on the BEC transition. We find this substantially shifts the transition temperature and so strengthens the bounds on cross section. In particular, for the well-motivated mass range of $~5-15\;{\rm GeV}$, we improve previous bounds by an order of magnitude.
\end{abstract}

\title{Effects of gravitational confinement on bosonic asymmetric dark matter in stars}
\maketitle
\section{Introduction}
\label{Intro}
While the existence of dark matter is at present well established, the nature of this dark matter is still a matter of considerable speculation \cite{Bertone05}. Constraints from precision cosmology \cite{Wang02} and observations of the Bullet Cluster \cite{Clowe06} give a picture of the generic phenomenology that dark matter models must satisfy. Direct detection experiments have also placed constraints on the interactions of the dark matter with baryons. Recently, attempts have been made to use astrophysical observations to further constrain the nature of the dark matter. However, such constraints tend to be strongly model dependent.

One particularly well-motivated class of dark matter models are known as asymmetric dark matter (ADM) models \cite{Nussinov85,DBKaplan92}. These models attempt to explain the similarity of dark matter and baryonic densities $\Omega_{\rm DM} \approx 5 \Omega_{\rm baryon}$ by postulating a mechanism for transferring the baryonic asymmetry of the standard model sector to the dark matter sector and visa versa. Such models favor a mass of the dark matter particles in the range of $~5-15\;{\rm GeV}$\cite{DEKaplan09}. The asymmetry in the dark sector implies that dark matter particles are stable and lack antiparticles with which to annihilate.

This stability allows for interesting bounds to be derived from the existence of old neutron stars, particularly for bosonic dark matter\cite{Kouvaris11,McDermott12}. The stability of ADM allows a neutron star to slowly accumulate a cloud of ADM particles by scattering them in such a way as to reduce their speed below the neutron star's escape speed. Once captured in the neutron star's gravitational field, repeated scattering between the star and the ADM particles allows the ADM to thermalize with the star. Once a sufficient amount of dark matter has accumulated, the ADM cloud can become self-gravitating and collapse into a black hole. This black hole then consumes the host neutron star. Bosonic ADM can undergo a phase transition to a Bose-Einstein condensate (BEC) state. The BEC, having higher density than a thermal gas cloud, considerably increases the strength of the bounds obtained by this method.

However, all analyses of this situation, to date, have treated the thermal ADM cloud as homogeneous. In this note, we consider the shortcomings of this approximation for a cloud of non-interacting bosons with a radius much smaller than that of the neutron star. Taking account of the effects of the confining gravitational potential improves the strength of the bounds for all masses up to $15\;\rm{GeV}$. The improvement is particularly strong, a factor of an order of magnitude, in the well-motivated mass range of $\approx 5-15\;{\rm GeV}$.
\section{Effects of Confinement}
\label{results}
The well-known newtonian theory of gravitation in a uniform density sphere suggests that the ADM cloud within the neutron star should experience a harmonic potential energy given by $V(r) = (2\pi G \rho m/3) r^2$. Calculations\cite{Bagnato87}, subsequently verified by experiments with dilute alkali gases\cite{Ensher96}, show the density of states for a gas trapped in a harmonic potential is markedly different from that of a homogeneous gas. Specifically, for a gas confined in an isotropic harmonic potential the density of states, $g(\epsilon)$, at energy $\epsilon$ is given by $g(\epsilon) = \epsilon^2/(2(\hbar \omega)^3)$, where $\omega$ is the angular frequency of the harmonic oscillator potential. This may be compared to the homogeneous gas result, $g(\epsilon) = \sqrt{\epsilon}(2 \pi V)(2m/h^2)^{3/2}$ , where $m$ is the dark matter particle mass and $V$ is the volume occupied by the gas.

For a fixed temperature, $T$, the critical number, $N_{\rm c}$, in the harmonic trap for which a BEC begins to form is $N_{\rm c} = \zeta(3)(kT/\hbar \omega)^3$, with $\zeta(x)$ the Riemann zeta function and $\zeta(3)\approx 1.2$. Using the newtonian potential above we find a critical number
\[
N_{\rm sho} = \zeta(3) \left({kT \over \hbar \sqrt{4\pi G \rho /3}} \right)^3
\]
Comparing this to the value, $N_{\rm hom}$, found using the homogeneous gas in a volume determined by gravitational binding in\cite{Kouvaris11,McDermott12} we see the same temperature scaling but a larger prefactor:
\[
{N_{\rm hom} \over N_{\rm sho}} = \sqrt{6 \over \pi} {\zeta(3/2) \over \zeta(3)} = 3.0.
\]

Using the newtonian potential for a relativistic system such as a neutron star, particularly near the core, misses important effects. We therefore derive the metric near the center of the star and from it derive an effective gravitational potential energy. To do so, we model the star as spherically symmetric with a metric of the form
\[ ds^2 = -e^{2\Phi} dt^2+e^{2\lambda}dr^2+r^2 d\Omega,\ e^{2\lambda} \equiv \left( 1-{2G m(r) \over r}\right) ^{-1}
\]
where $m(r)$ is the mass contained within a sphere of radial coordinate $r$. For the temperatures of interest, $T\approx 10^5-10^7\;{\rm K}$ at the neutron star core, the thermal ADM cloud should be localized within a sphere of radius roughly $1\;{\rm m}$ around the center of the neutron star. Thus it is reasonable to treat the density as uniform in the region of interest. This gives us $m(r) = 4\pi\rho r^3/3$, making $g_{rr}=1-8\pi r^2\rho /3$. Using $\rho = 1.5\times10^{15}\;{\rm g/cm^3}$ shows that for $r\le 1\;{\rm m}$, $|g_{rr}-1|\le 10^{-8}$. Thus, we can neglect the curvature of the spatial components of the metric and treat the radial coordinate as the proper length, simplifying the analysis. This allows us to treat the system via an effective newtonian potential, $\Phi$. Using the Oppenheimer-Volkoff equation\cite{Shapiro83} we find
\begin{align*}
\frac{d\Phi}{dr} &=\frac{Gm}{r^2} \left(1+{4 \pi P r^3 \over m} \right)\left(1-{2Gm \over r} \right)^{-1}\\
&= {4\pi\over 3}G\rho r \left(1+3{P\over \rho} \right),
\end{align*}
where P is the pressure at the star's core.

From this expression, we see that the newtonian analysis can be carried over by simply replacing $\rho$ with $\rho +3P$. The pressure depends sensitively on the neutron star equation of state. Recently, observations of x-ray emission from neutron stars have progressed to the point where good approximations for this equation of state can be had\cite{Ozel10,Steiner12}. The core density used above corresponds to $\rho \approx 0.9\;{\rm GeV/fm^3}$ with a pressure of $P \approx 0.3\;{\rm GeV/fm^3}$. The relativistic analysis therefore gives a gravitational potential roughly twice as strong as the newtonian analysis. This further decreases the critical number for the condensation transition. Our final result for the critical number is thus
\[
N_{\rm sho} = \zeta(3) \left({kT \over \hbar \sqrt{4\pi G (\rho+3P) /3}} \right)^3.
\]

For the mass range $~5-15\;{\rm GeV}$ the constraints on dark matter/nucleon scattering are dominated by the critical number. As such, using the corrected expression above strengthens the bounds in this range by roughly a factor of 8 as compared to those from ref \cite{McDermott12}. Further refinements to the neutron star equation of state are anticipated in the near future, based on discrepancies between complementary techniques for deriving it from observation\cite{Steiner12,Guillot13} as well as the accumulations of further data. Bounds on dark matter cross-sections may be easily updated in the light of such new information, based on the above results.

\section{Acknowledgements}
 I thank A. Nelson and S. Reddy for helpful discussions and readings of the manuscript. This work was supported by a NIST Precision Measurement Grant and the Department of Energy under grant DE-FG02-96ER40956.


\begin{thebibliography}{14}%
\makeatletter
\providecommand \@ifxundefined [1]{%
 \@ifx{#1\undefined}
}%
\providecommand \@ifnum [1]{%
 \ifnum #1\expandafter \@firstoftwo
 \else \expandafter \@secondoftwo
 \fi
}%
\providecommand \@ifx [1]{%
 \ifx #1\expandafter \@firstoftwo
 \else \expandafter \@secondoftwo
 \fi
}%
\providecommand \natexlab [1]{#1}%
\providecommand \enquote  [1]{``#1''}%
\providecommand \bibnamefont  [1]{#1}%
\providecommand \bibfnamefont [1]{#1}%
\providecommand \citenamefont [1]{#1}%
\providecommand \href@noop [0]{\@secondoftwo}%
\providecommand \href [0]{\begingroup \@sanitize@url \@href}%
\providecommand \@href[1]{\@@startlink{#1}\@@href}%
\providecommand \@@href[1]{\endgroup#1\@@endlink}%
\providecommand \@sanitize@url [0]{\catcode `\\12\catcode `\$12\catcode
  `\&12\catcode `\#12\catcode `\^12\catcode `\_12\catcode `\%12\relax}%
\providecommand \@@startlink[1]{}%
\providecommand \@@endlink[0]{}%
\providecommand \url  [0]{\begingroup\@sanitize@url \@url }%
\providecommand \@url [1]{\endgroup\@href {#1}{\urlprefix }}%
\providecommand \urlprefix  [0]{URL }%
\providecommand \Eprint [0]{\href }%
\providecommand \doibase [0]{http://dx.doi.org/}%
\providecommand \selectlanguage [0]{\@gobble}%
\providecommand \bibinfo  [0]{\@secondoftwo}%
\providecommand \bibfield  [0]{\@secondoftwo}%
\providecommand \translation [1]{[#1]}%
\providecommand \BibitemOpen [0]{}%
\providecommand \bibitemStop [0]{}%
\providecommand \bibitemNoStop [0]{.\EOS\space}%
\providecommand \EOS [0]{\spacefactor3000\relax}%
\providecommand \BibitemShut  [1]{\csname bibitem#1\endcsname}%
\let\auto@bib@innerbib\@empty
\bibitem [{\citenamefont {Bertone}\ \emph {et~al.}(2005)\citenamefont
  {Bertone}, \citenamefont {Hooper},\ and\ \citenamefont {Silk}}]{Bertone05}%
  \BibitemOpen
  \bibfield  {author} {\bibinfo {author} {\bibfnamefont {G.}~\bibnamefont
  {Bertone}}, \bibinfo {author} {\bibfnamefont {D.}~\bibnamefont {Hooper}}, \
  and\ \bibinfo {author} {\bibfnamefont {J.}~\bibnamefont {Silk}},\ }\href
  {\doibase 10.1016/j.physrep.2004.08.031} {\bibfield  {journal} {\bibinfo
  {journal} {Physics Reports}\ }\textbf {\bibinfo {volume} {405}},\ \bibinfo
  {pages} {279 } (\bibinfo {year} {2005})},\ \Eprint
  {http://arxiv.org/abs/hep-ph/0404175} {hep-ph/0404175} \BibitemShut {NoStop}%
\bibitem [{\citenamefont {Wang}\ \emph {et~al.}(2002)\citenamefont {Wang},
  \citenamefont {Tegmark},\ and\ \citenamefont {Zaldarriaga}}]{Wang02}%
  \BibitemOpen
  \bibfield  {author} {\bibinfo {author} {\bibfnamefont {X.}~\bibnamefont
  {Wang}}, \bibinfo {author} {\bibfnamefont {M.}~\bibnamefont {Tegmark}}, \
  and\ \bibinfo {author} {\bibfnamefont {M.}~\bibnamefont {Zaldarriaga}},\
  }\href {\doibase 10.1103/PhysRevD.65.123001} {\bibfield  {journal} {\bibinfo
  {journal} {Phys. Rev. D}\ }\textbf {\bibinfo {volume} {65}},\ \bibinfo
  {pages} {123001} (\bibinfo {year} {2002})},\ \Eprint
  {http://arxiv.org/abs/astro-ph/0105091} {astro-ph/0105091} \BibitemShut
  {NoStop}%
\bibitem [{\citenamefont {Clowe}\ \emph {et~al.}(2006)\citenamefont {Clowe},
  \citenamefont {Bradač}, \citenamefont {Gonzalez}, \citenamefont
  {Markevitch}, \citenamefont {Randall}, \citenamefont {Jones},\ and\
  \citenamefont {Zaritsky}}]{Clowe06}%
  \BibitemOpen
  \bibfield  {author} {\bibinfo {author} {\bibfnamefont {D.}~\bibnamefont
  {Clowe}}, \bibinfo {author} {\bibfnamefont {M.}~\bibnamefont {Bradač}},
  \bibinfo {author} {\bibfnamefont {A.~H.}\ \bibnamefont {Gonzalez}}, \bibinfo
  {author} {\bibfnamefont {M.}~\bibnamefont {Markevitch}}, \bibinfo {author}
  {\bibfnamefont {S.~W.}\ \bibnamefont {Randall}}, \bibinfo {author}
  {\bibfnamefont {C.}~\bibnamefont {Jones}}, \ and\ \bibinfo {author}
  {\bibfnamefont {D.}~\bibnamefont {Zaritsky}},\ }\href
  {http://stacks.iop.org/1538-4357/648/i=2/a=L109} {\bibfield  {journal}
  {\bibinfo  {journal} {The Astrophysical Journal Letters}\ }\textbf {\bibinfo
  {volume} {648}},\ \bibinfo {pages} {L109} (\bibinfo {year} {2006})},\ \Eprint
  {http://arxiv.org/abs/astro-ph/0608407} {astro-ph/0608407} \BibitemShut
  {NoStop}%
\bibitem [{\citenamefont {Nussinov}(1985)}]{Nussinov85}%
  \BibitemOpen
  \bibfield  {author} {\bibinfo {author} {\bibfnamefont {S.}~\bibnamefont
  {Nussinov}},\ }\href {\doibase 10.1016/0370-2693(85)90689-6} {\bibfield
  {journal} {\bibinfo  {journal} {Physics Letters B}\ }\textbf {\bibinfo
  {volume} {165}},\ \bibinfo {pages} {55 } (\bibinfo {year}
  {1985})}\BibitemShut {NoStop}%
\bibitem [{\citenamefont {Kaplan}(1992)}]{DBKaplan92}%
  \BibitemOpen
  \bibfield  {author} {\bibinfo {author} {\bibfnamefont {D.~B.}\ \bibnamefont
  {Kaplan}},\ }\href {\doibase 10.1103/PhysRevLett.68.741} {\bibfield
  {journal} {\bibinfo  {journal} {Phys. Rev. Lett.}\ }\textbf {\bibinfo
  {volume} {68}},\ \bibinfo {pages} {741} (\bibinfo {year} {1992})}\BibitemShut
  {NoStop}%
\bibitem [{\citenamefont {Kaplan}\ \emph {et~al.}(2009)\citenamefont {Kaplan},
  \citenamefont {Luty},\ and\ \citenamefont {Zurek}}]{DEKaplan09}%
  \BibitemOpen
  \bibfield  {author} {\bibinfo {author} {\bibfnamefont {D.~E.}\ \bibnamefont
  {Kaplan}}, \bibinfo {author} {\bibfnamefont {M.~A.}\ \bibnamefont {Luty}}, \
  and\ \bibinfo {author} {\bibfnamefont {K.~M.}\ \bibnamefont {Zurek}},\ }\href
  {\doibase 10.1103/PhysRevD.79.115016} {\bibfield  {journal} {\bibinfo
  {journal} {Phys. Rev. D}\ }\textbf {\bibinfo {volume} {79}},\ \bibinfo
  {pages} {115016} (\bibinfo {year} {2009})},\ \Eprint
  {http://arxiv.org/abs/0901.4117} {arXiv:0901.4117 [hep-ph]} \BibitemShut
  {NoStop}%
\bibitem [{\citenamefont {Kouvaris}\ and\ \citenamefont
  {Tinyakov}(2011)}]{Kouvaris11}%
  \BibitemOpen
  \bibfield  {author} {\bibinfo {author} {\bibfnamefont {C.}~\bibnamefont
  {Kouvaris}}\ and\ \bibinfo {author} {\bibfnamefont {P.}~\bibnamefont
  {Tinyakov}},\ }\href {\doibase 10.1103/PhysRevLett.107.091301} {\bibfield
  {journal} {\bibinfo  {journal} {Phys. Rev. Lett.}\ }\textbf {\bibinfo
  {volume} {107}},\ \bibinfo {pages} {091301} (\bibinfo {year} {2011})},\
  \Eprint {http://arxiv.org/abs/1104.0382} {arXiv:1104.0382 [astro-ph]}
  \BibitemShut {NoStop}%
\bibitem [{\citenamefont {McDermott}\ \emph {et~al.}(2012)\citenamefont
  {McDermott}, \citenamefont {Yu},\ and\ \citenamefont {Zurek}}]{McDermott12}%
  \BibitemOpen
  \bibfield  {author} {\bibinfo {author} {\bibfnamefont {S.~D.}\ \bibnamefont
  {McDermott}}, \bibinfo {author} {\bibfnamefont {H.-B.}\ \bibnamefont {Yu}}, \
  and\ \bibinfo {author} {\bibfnamefont {K.~M.}\ \bibnamefont {Zurek}},\ }\href
  {\doibase 10.1103/PhysRevD.85.023519} {\bibfield  {journal} {\bibinfo
  {journal} {Phys. Rev. D}\ }\textbf {\bibinfo {volume} {85}},\ \bibinfo
  {pages} {023519} (\bibinfo {year} {2012})},\ \Eprint
  {http://arxiv.org/abs/1103.5472} {arXiv:1103.5472 [hep-ph]} \BibitemShut
  {NoStop}%
\bibitem [{\citenamefont {Bagnato}\ \emph {et~al.}(1987)\citenamefont
  {Bagnato}, \citenamefont {Pritchard},\ and\ \citenamefont
  {Kleppner}}]{Bagnato87}%
  \BibitemOpen
  \bibfield  {author} {\bibinfo {author} {\bibfnamefont {V.}~\bibnamefont
  {Bagnato}}, \bibinfo {author} {\bibfnamefont {D.~E.}\ \bibnamefont
  {Pritchard}}, \ and\ \bibinfo {author} {\bibfnamefont {D.}~\bibnamefont
  {Kleppner}},\ }\href {\doibase 10.1103/PhysRevA.35.4354} {\bibfield
  {journal} {\bibinfo  {journal} {Phys. Rev. A}\ }\textbf {\bibinfo {volume}
  {35}},\ \bibinfo {pages} {4354} (\bibinfo {year} {1987})}\BibitemShut
  {NoStop}%
\bibitem [{\citenamefont {Ensher}\ \emph {et~al.}(1996)\citenamefont {Ensher},
  \citenamefont {Jin}, \citenamefont {Matthews}, \citenamefont {Wieman},\ and\
  \citenamefont {Cornell}}]{Ensher96}%
  \BibitemOpen
  \bibfield  {author} {\bibinfo {author} {\bibfnamefont {J.~R.}\ \bibnamefont
  {Ensher}}, \bibinfo {author} {\bibfnamefont {D.~S.}\ \bibnamefont {Jin}},
  \bibinfo {author} {\bibfnamefont {M.~R.}\ \bibnamefont {Matthews}}, \bibinfo
  {author} {\bibfnamefont {C.~E.}\ \bibnamefont {Wieman}}, \ and\ \bibinfo
  {author} {\bibfnamefont {E.~A.}\ \bibnamefont {Cornell}},\ }\href {\doibase
  10.1103/PhysRevLett.77.4984} {\bibfield  {journal} {\bibinfo  {journal}
  {Phys. Rev. Lett.}\ }\textbf {\bibinfo {volume} {77}},\ \bibinfo {pages}
  {4984} (\bibinfo {year} {1996})}\BibitemShut {NoStop}%
\bibitem [{\citenamefont {Shapiro}\ and\ \citenamefont
  {Teukolsky}(1983)}]{Shapiro83}%
  \BibitemOpen
  \bibfield  {author} {\bibinfo {author} {\bibfnamefont {S.~L.}\ \bibnamefont
  {Shapiro}}\ and\ \bibinfo {author} {\bibfnamefont {S.~A.}\ \bibnamefont
  {Teukolsky}},\ }\href@noop {} {\emph {\bibinfo {title} {Black Holes, White
  Dwarfs, and Neutron Stars}}}\ (\bibinfo  {publisher} {Wiley Interscience},\
  \bibinfo {address} {New York, NY},\ \bibinfo {year} {1983})\BibitemShut
  {NoStop}%
\bibitem [{\citenamefont {\"Ozel}\ \emph {et~al.}(2010)\citenamefont {\"Ozel},
  \citenamefont {Baym},\ and\ \citenamefont {G\"uver}}]{Ozel10}%
  \BibitemOpen
  \bibfield  {author} {\bibinfo {author} {\bibfnamefont {F.}~\bibnamefont
  {\"Ozel}}, \bibinfo {author} {\bibfnamefont {G.}~\bibnamefont {Baym}}, \ and\
  \bibinfo {author} {\bibfnamefont {T.}~\bibnamefont {G\"uver}},\ }\href
  {\doibase 10.1103/PhysRevD.82.101301} {\bibfield  {journal} {\bibinfo
  {journal} {Phys. Rev. D}\ }\textbf {\bibinfo {volume} {82}},\ \bibinfo
  {pages} {101301} (\bibinfo {year} {2010})},\ \Eprint
  {http://arxiv.org/abs/1002.3153} {arXiv:1002.3153 [astro-ph]} \BibitemShut
  {NoStop}%
\bibitem [{\citenamefont {Andrew W.~Steiner}(2012)}]{Steiner12}%
  \BibitemOpen
  \bibfield  {author} {\bibinfo {author} {\bibfnamefont {E.~F.~B.}\
  \bibnamefont {Andrew W.~Steiner}, \bibfnamefont {James M.~Lattimer}},\
  }\href@noop {} {\  (\bibinfo {year} {2012})},\ \Eprint
  {http://arxiv.org/abs/1205.6871} {arXiv:1205.6871 [nucl-th]} \BibitemShut
  {NoStop}%
\bibitem [{\citenamefont {Guillot}\ \emph {et~al.}(2013)\citenamefont
  {Guillot}, \citenamefont {Servillat}, \citenamefont {Webb},\ and\
  \citenamefont {Rutledge}}]{Guillot13}%
  \BibitemOpen
  \bibfield  {author} {\bibinfo {author} {\bibfnamefont {S.}~\bibnamefont
  {Guillot}}, \bibinfo {author} {\bibfnamefont {M.}~\bibnamefont {Servillat}},
  \bibinfo {author} {\bibfnamefont {N.}~\bibnamefont {Webb}}, \ and\ \bibinfo
  {author} {\bibfnamefont {R.}~\bibnamefont {Rutledge}},\ }\href@noop {} {\
  (\bibinfo {year} {2013})},\ \Eprint {http://arxiv.org/abs/1302.0023}
  {arXiv:1302.0023 [nucl-th]} \BibitemShut {NoStop}%
\end{thebibliography}
\end{document}